\documentclass[aps,prd,reprint,showpacs,longbibliography,groupedaddress,titlepage,
 amsmath,amssymb,
]{revtex4-1}
\usepackage[utf8]{inputenc}
\usepackage{tensor}
\usepackage{undertilde}
\usepackage{accents}
\usepackage{hyperref}
\usepackage{graphicx}
\usepackage{dcolumn}
\usepackage{bm}
\hypersetup{colorlinks=true,linkcolor=blue,urlcolor=blue,citecolor=blue}

\usepackage[none]{hyphenat}

\usepackage[protrusion=true,tracking=true,kerning=true,final,spacing=true,stretch=10,shrink=10]{microtype}

\begin{document}


\title{Alternative derivation of Krasnov's action for general relativity}

\author{Mariano Celada}
\email[]{mcelada@fis.cinvestav.mx}
\affiliation{Departamento de F\'{\i}sica, Cinvestav, Instituto Polit\'ecnico Nacional 2508, San Pedro Zacatenco,
	07360, Gustavo A. Madero, Ciudad de M\'exico, M\'exico}

\author{Diego Gonz\'alez}
\email[]{dgonzalez@fis.cinvestav.mx}
\affiliation{Departamento de F\'{\i}sica, Cinvestav, Instituto Polit\'ecnico Nacional 2508, San Pedro Zacatenco,
	07360, Gustavo A. Madero, Ciudad de M\'exico, M\'exico}

\author{Merced Montesinos} 
\email[]{merced@fis.cinvestav.mx}
\affiliation{Departamento de F\'{\i}sica, Cinvestav, Instituto Polit\'ecnico Nacional 2508, San Pedro Zacatenco, 
	07360, Gustavo A. Madero, 
	Ciudad de M\'exico, M\'exico}

\date{\today}

\begin{abstract}
Starting from Plebanski's action for general relativity with cosmological constant, we show that by integrating out all the auxiliary fields Krasnov's action immediately emerges. We also perform the Hamiltonian analysis of the latter and show that the constraints are those of the Ashtekar formalism.
\end{abstract}

\maketitle


\section{Introduction}\label{intro}

Krasnov's action principle is a pure connection formulation for complex general relativity with a nonvanishing cosmological constant. It was proposed in Ref.~\cite{krasnprl} and it was shown there that this action principle is what remains once one integrates out some of the auxiliary fields involved in Plebanski's action. However, this approach becomes a little tricky in the last step, when the field $\Psi$ must be eliminated from the action, because it requires going to a particular basis where the curvature matrix becomes diagonal and then certain conditions on its eigenvalues must be imposed. On the other hand, we show here that it is possible to obtain the same pure connection action in a cleaner fashion. The strategy we follow is simpler and has as the initial starting point the Plebanski formulation where all the relevant fields and constraints are encompassed. The key point of our method consists in explicitly adding to the Plebanski action (through a Lagrange multiplier) the condition that the field $\Psi$ has to be traceless (in~\cite{krasnprl} the field $\Psi$ is traceless by itself), and this, together with some facts about square roots of matrices, allows us to go around the aforementioned approach. Thus, by following a systematic procedure in which we integrate out all the auxiliary fields step by step, we finally arrive at the Krasnov formulation. Since a detailed Hamiltonian analysis of this formulation has not been performed yet, we carry out such an analysis here and find that the associated phase space and constraints are the same as those of the Ashtekar formalism.

The achievement of a formulation of gravity as a diffeomorphism-invariant pure connection theory has been one of the most tackled problems through the years (notice that the first attempt to formulate general relativity as such is due to Eddington~\cite{Eddington}). For instance, this formulation might help in the search for a unified description of general relativity and the gauge theories describing the standard model of particle physics. Since initially general relativity was formulated as a metric theory, the introduction of 2-forms as fundamental fields established the first attempt towards formulating gravity as a gauge theory~\cite{pleb1977118}, although new fields were introduced in order to recover the metric at the end; the metric itself thus became a derived object, and it was later realized that the Urbantke metric~\cite{urba198425} constructed from the 2-forms was the right choice as the spacetime metric.

Later, at the end of the eighties the first almost pure connection formulation for general relativity was given~\cite{capo198963}: the Capovilla-Dell-Jacobson (CDJ) formulation showed how to write general relativity with vanishing cosmological constant as a gauge theory depending only on a SO(3,$\mathbb{C}$) connection and a Lagrange multiplier. The case of a nonvanishing cosmological constant turned out to be harder than the case $\Lambda=0$, but it was finally attained~\cite{Peld8-10-005,capo199207}. Notice that another kind of generalization of the CDJ action admitting an infinite number of parameters, one of which could be identified with the cosmological constant, was also reported in~\cite{Bengt199155}.

At around the same time, the discovery of the Ashtekar formalism~\cite{AshPRD.36.1587} for describing the phase space of general relativity drew the researchers' attention towards the development of a nonperturbative quantum theory of gravity. Although Ashtekar variables were originally obtained through a canonical transformation performed on the ADM variables, they naturally arise from the Hamiltonian analysis of the Plebanski action~\cite{capo-cqg8-41}. The Hamiltonian formulation of the CDJ action with $\Lambda=0$ also leads to a phase space described by the variables and constraints of the Ashtekar formalism (the same holds for the intermediate step towards it in the case of $\Lambda\neq0$~\cite{Dadhich-8-3-002}), which shows again the equivalence of the CDJ formulation with general relativity, and even the generalizations proposed in~\cite{Bengt199155} have the same number of physical degrees (DOF) of freedom of gravity when they are written in terms of such variables.

Thus, the metric was practically given up as the appropriate variable for describing the phase space of general relativity, and the gauge formulations of gravity soon became the starting point of the emerging (canonical) quantization approaches of gravity (for instance, the Ashtekar formulation established the foundations for the loop quantum gravity approach~\cite{RovellilR}, whereas the Plebanski formulation was taken as the starting point of the spinfoam models~\cite{PerezLR}), and in fact all the efforts turned towards the successful accomplishment of this long-dreamed quantum theory. However, none of the previous formulations of gravity was a pure connection one (although the closest ones differed from it just by a nondynamical variable).

Before the pure connection formulation of gravity was discovered, somehow a more general class of diffeomorphism-invariant gauge theories [depending only on a SO(3,$\mathbb{C}$) gauge connection] were conceived in Ref.~\cite{Krasn81.084026}; this class of theories, whose action principle takes the form $S[A]=\int d^4 x\ f(F^i\wedge F^j)$, with $f$ being a holomorphic function homogeneous of degree one and gauge invariant, also propagates two degrees of freedom~\cite{Krasn84.024034} (notice, however,  that the Pontryagin term, which is topological, also belongs to this class of theories). In fact, it turned out that the Krasnov's action describing general relativity with cosmological constant was a particular member of this class of theories, the one for which $f\sim(\text{Tr}\sqrt{F^i\wedge F^j})^2$, and then the search for a pure connection formulation of general relativity finally came to an end.

\section{From Plebanski to Krasnov}

We start from the Plebanski formulation for general relativity with cosmological constant, which is given by~\cite{pleb1977118,capo-cqg8-41,capo199207}

\begin{eqnarray}
S[A, \Sigma,\Psi,\rho]=\int &\biggl[ &\Sigma_i \wedge F^i\notag\\ &&\hspace{-20mm}-\frac{1}{2} \left(\Psi_{ij}+\frac{1}{3}\Lambda \delta_{ij}  \right)  \Sigma^i \wedge \Sigma^j -\rho {\rm Tr}\Psi  \biggr],   \label{plebanAct}
\end{eqnarray}

\noindent where $\Sigma^i$ are three $\mathfrak{so}(3,\mathbb{C})$-valued 2-forms, $F^i=d A^i + (1/2) \tensor{\varepsilon}{^i_{jk}}  A^j  \wedge A^k$ is the curvature of the $SO(3,\mathbb{C})$ gauge connection $A^i$,  $\Psi$ is a $3\times 3$ complex symmetric matrix, $\rho$ is a Lagrange multiplier and $\Lambda$ is the cosmological constant. The indices $i,j,\ldots =1,2,3$ are raised and lowered with the Kronecker delta $\delta^{ij}$ and ${\varepsilon}_{ijk}$ is the Levi-Civit\`a symbol  (${\varepsilon}_{123}=+1$). The action (\ref{plebanAct}) reduces to the self-dual Palatini action (with cosmological constant) when the constraint imposed on the field $\Sigma$ is solved and the solution is put back into the action~\cite{samuel,Jacobson198739,Jacobsoncqg5}, but in order to make contact with Krasnov's action we follow another approach.  First of all, the equation of motion corresponding to $\Sigma^i$ is

\begin{equation}\label{plebeqf}
	F^i=\left(\Psi^i{}_j+\frac{\Lambda}{3}\delta^i_j\right) \Sigma^j.
\end{equation}

\noindent From this expression we want to express the 2-form field $\Sigma$ in terms of the remaining fields and put it back into the action (\ref{plebanAct}). In such a way, we obtain an action principle which is classically equivalent to the initial one~\cite{teitel}, but with fewer variables. Let us define the 3$\times$3 symmetric matrix $X_{ij}:=\Psi_{ij}+(\Lambda/3)\delta_{ij}$ and suppose that it is nonsingular~\cite{capo1991859,capo199207,krasnprl}. Then, Eq. (\ref{plebeqf}) can be rewritten as

\begin{equation}\label{plebeqsig}
	\Sigma^i=\tensor{(X^{-1})}{^i_j}F^j.
\end{equation}

\noindent It is important to stress that the invertibility of $X$ is necessary in order to solve Eq. (\ref{plebeqf}) for $\Sigma$, and this implies that the case when $X$ is singular is excluded from our approach. Substituting Eq. (\ref{plebeqsig}) into Eq. (\ref{plebanAct}) yields

\begin{equation}\label{plebAPR}
S[A,\Psi,\rho]=\int \left[ \frac{1}{2} (X^{-1})_{ij}F^i\wedge F^j-\rho {\rm Tr}\Psi  \right].
\end{equation}

The next step consists in integrating out the field $\Psi$ from the action principle (\ref{plebAPR}). We first write (\ref{plebAPR}) as

\begin{equation}\label{plebAPR1}
	S[A,\Psi,\rho]=\int d^4x \left[ \frac{1}{2} (X^{-1})_{ij}\tilde{M}^{ij} - {\tilde \rho} {\rm Tr}\Psi\right],	
\end{equation}

\noindent where we have defined $F^i\wedge F^j=:\tilde{M}^{ij}d^4x$ and $\rho=:\tilde{\rho}\ d^4x$. The variation of (\ref{plebAPR1}) with respect to $\Psi$ leads to

\begin{equation}\label{eqPsi}
	\tilde{M}+2\tilde{\rho}X^2=0,
\end{equation}

\noindent which now must be solved for $\Psi$. Since $X$ is nonsingular, then, from (\ref{eqPsi}), $\tilde{M}$ is also nonsingular. Besides, this matrix equation says that $X$ is essentially the square root of $\tilde{M}$. Although square roots of matrices do not always exist, the fact that $\tilde{M}$ is invertible guarantees the existence of a square root of $\tilde{M}$~\cite{horn1994topics} (which in this case is also symmetric), albeit it is not unique in general. Therefore, Eq. (\ref{eqPsi}) implies

\begin{equation}\label{eqPsisolv}
	X(A,\rho)=\frac{\text{i}}{\sqrt{2}} \tilde{\rho}^{-1/2} \tilde{M}^{1/2},
\end{equation}

\noindent where $\tilde{M}=\tilde{M}^{1/2} \tilde{M}^{1/2}$, and we have absorbed all the arbitrariness carried by the square root (like the choice of the branch) in the definition of $\tilde{M}^{1/2}$.  We point out that the expression (\ref{eqPsisolv}) involves a choice of a particular square root of $\tilde{M}$, but the procedure we follow is independent of the specific root chosen, that is, any (symmetric) square root of $\tilde{M}$ can be taken. Inserting (\ref{eqPsisolv}) into the action principle (\ref{plebAPR1}) yields

\begin{equation}\label{plebAR}
	S[A,\rho]=-\int d^4x \left( \sqrt{2} \text{i} \tilde{\rho}^{1/2} {\text{Tr}}\tilde{M}^{1/2}-\Lambda\tilde{\rho} \right).
\end{equation}

Notice that here (\ref{eqPsisolv}) and (\ref{plebAR}) are results, not hypotheses as in Ref.~\cite{capo1991859}. The action principle (\ref{plebAR}) describes general relativity for both vanishing and nonvanishing cosmological constant. In fact, for $\Lambda=0$ this action constitutes an intermediate step in the road towards the CDJ formulation~\cite{capo1991859}.

We are just one step behind Krasnov's action; all we need to do is to integrate out the field $\rho$ from the action (\ref{plebAR}). The equation of motion for $\rho$ from (\ref{plebAR}) is

\begin{equation}\label{eqrho}
	\frac{\text{i}}{\sqrt{2}} \tilde{\rho}^{-1/2} {\text{Tr}}\tilde{M}^{1/2}-\Lambda=0.
\end{equation}

\noindent Note that this equation can be solved for $\tilde{\rho}$ only if $\Lambda\neq 0$. In this case, $\tilde{\rho}$ takes the form

\begin{equation}\label{eqrhosolv}
	\tilde{\rho} = - \frac{1}{2 \Lambda^2}  ( {\rm Tr}\tilde{M}^{1/2})^2.
\end{equation}

\noindent By substituting this expression back into Eq. (\ref{plebAR}) we finally arrive at the action principle

\begin{equation}\label{krasnaction}
	S[A]=\frac{1}{2 \Lambda} \int d^4x ({\rm Tr}\tilde{M}^{1/2})^2,
\end{equation}

\noindent which constitutes the Krasnov formulation of general relativity with nonvanishing cosmological constant~\cite{krasnprl}.

\section{Hamiltonian analysis of Krasnov's action}

In this section we perform the canonical analysis of the action principle (\ref{krasnaction}), which sets up the first step towards a canonical quantization of this formulation. For such a purpose, we perform the 3+1 decomposition of the action (\ref{krasnaction}). We foliate the spacetime by 3-manifolds $\Omega_t$ at a constant global time function $t$, so that the spacetime has the topology $\mathbb{R}\times\Omega$, where $\Omega$ is a spacial compact 3-manifold without a boundary. We refer to the time component as the 0-component, and denote the spacial indices by $a,\ b,\ldots =1,2,3$. From the definition of the matrix $\tilde{M}$ [see the line after Eq. (\ref{plebAPR1})], we obtain 

\begin{eqnarray}
\tilde{M}^{ij}&=&\frac{1}{4}\tilde{\eta}^{\mu\nu\lambda\sigma}\tensor{F}{_{\mu\nu}^i}\tensor{F}{_{\lambda\sigma}^j}\nonumber\\
&=&\frac{1}{2}\tilde{\eta}^{abc}\left(\tensor{F}{_{0a}^i}\tensor{F}{_{bc}^j}+\tensor{F}{_{0a}^j}\tensor{F}{_{bc}^i}\right),\label{mdec}
\end{eqnarray}

\noindent where $\tilde{\eta}^{\mu\nu\lambda\sigma}$ is a totally antisymmetric tensor density of weight 1 ($\tilde{\eta}^{0123}=+1$) and $\tilde{\eta}^{abc}:=\tilde{\eta}^{0abc}$. We denote by $\tilde{M}^{-1/2}$ to the inverse of $\tilde{M}^{1/2}$; multiplying (\ref{mdec}) by $(\tilde{M}^{-1/2})_{ij}$ yields

\begin{equation}\label{trsqM}
\text{Tr}\tilde{M}^{1/2}=\tilde{\eta}^{abc}\tensor{F}{_{0a}^i}(\tilde{M}^{-1/2})_{ij}\tensor{F}{_{bc}^j}.
\end{equation}

\noindent This expression implies

\begin{equation}\label{Krasd}
\frac{1}{2\Lambda}\left(\text{Tr}\tilde{M}^{1/2}\right)^{2}=\tensor{F}{_{0a}^i}\tensor{\tilde{\Pi}}{^a_i},
\end{equation}

\noindent with $\tensor{\tilde{\Pi}}{^a_i}$ defined by 

\begin{equation}\label{momenta}
\tensor{\tilde{\Pi}}{^a_i}:=\frac{1}{2\Lambda}\text{Tr}\tilde{M}^{1/2}(\tilde{M}^{-1/2})_{ij}\tilde{\eta}^{abc}\tensor{F}{_{bc}^j}.
\end{equation}

Now, by using the expression for the components of the curvature, namely $\tensor{F}{_{0a}^i}=\tensor{\dot{A}}{_a^i}-\partial_{a}\tensor{A}{_0^i}+\tensor{\varepsilon}{^i_{jk}}\tensor{A}{_0^j}\tensor{A}{_a^k}$ (a dot over a variable means a time derivative of such a variable), Eq. (\ref{Krasd}) reads

\begin{equation}\label{Krasd1}
\hspace{-2mm}\frac{1}{2\Lambda}\left(\text{Tr}\tilde{M}^{1/2}\right)^{2}=\tensor{\tilde{\Pi}}{^a_i}\tensor{\dot{A}}{_a^i}+\tensor{A}{_0^i}D_a\tensor{\tilde{\Pi}}{^a_i}-\partial_{a}(\tensor{\tilde{\Pi}}{^a_i}\tensor{A}{_0^i}).
\end{equation}

\noindent By inserting (\ref{Krasd1}) into the action (\ref{krasnaction}) we obtain

\begin{equation}\label{Krasnov1}
S[A]=\int_{\mathbb{R}}dt\int_{\Omega}d^3x\left(\tensor{\tilde{\Pi}}{^a_i}\tensor{\dot{A}}{_a^i}+\tensor{A}{_0^i}D_a\tensor{\tilde{\Pi}}{^a_i}\right),
\end{equation}

\noindent where the total derivative in (\ref{Krasd1}) vanishes since $\Omega$ has no boundary, and $D_a$ is the SO(3,$\mathbb{C}$)-covariant derivative. From this expression we identify the canonical pair $(\tensor{A}{_a^i},\tensor{\tilde{\Pi}}{^a_i})$ whose fundamental Poisson bracket satisfies $\left\{\tensor{A}{_a^i}(x),\tensor{\tilde{\Pi}}{^b_j}(y)\right\}=\delta^b_a\delta^i_j\delta^3(x-y)$. Notice that $\tensor{A}{_0^i}$ appears linearly in the action (\ref{Krasnov1}), and so it plays the role of a Lagrange multiplier imposing the constraint

\begin{equation}\label{Gauss}
\tilde{\mathcal{G}}_i:=D_a\tensor{\tilde{\Pi}}{^a_i}\approx 0,
\end{equation}

\noindent which is the Gauss constraint that generates SO(3,$\mathbb{C}$) transformations. But this is not the end of the story, since there are more constraints coming from the definition of the canonical momenta. Indeed, from Eq. (\ref{momenta}) we find the following primary constraints,

\begin{eqnarray}
&\tilde{\mathcal{V}}_a:=\tensor{\tilde{\Pi}}{^b_i}\tensor{F}{_{ba}^i}\approx 0,\label{vector}\\
&\hspace{-6mm}\tilde{\tilde{\mathcal{H}}}:=\utilde{\eta}_{abc}\varepsilon_{ijk}\tilde{\Pi}{^{ai}}\tilde{\Pi}{^{bj}}\tilde{B}{^{ck}}-\frac{\Lambda}{3}\utilde{\eta}_{abc}\varepsilon_{ijk}\tilde{\Pi}{^{ai}}\tilde{\Pi}{^{bj}}\tilde{\Pi}{^{ck}}\approx 0\label{scalar},
\end{eqnarray}

\noindent where $\tilde{B}{^{ai}}:=(1/2)\tilde{\eta}^{abc}\tensor{F}{_{bc}^i}$. The expressions (\ref{vector})-(\ref{scalar}) define the vector and the scalar constraints, respectively. To include these constraints into the formalism, we introduce new Lagrange multipliers $N^a$ and $\utilde{N}$ so that the extended action reads

\begin{eqnarray}
& S[\tensor{A}{_a^i},\tensor{\tilde{\Pi}}{^a_i},\tensor{A}{_0^i},N^a,\utilde{N}]=\nonumber\\
&\int_{\mathbb{R}}dt\int_{\Omega}d^3x\left(\tensor{\tilde{\Pi}}{^a_i}\tensor{\dot{A}}{_a^i}+\tensor{A}{_0^i}\tilde{\mathcal{G}}_i+N^a\tilde{\mathcal{V}}_a+\utilde{N}\tilde{\tilde{\mathcal{H}}}\right).\label{Kras2}
\end{eqnarray}

Since the constraints are first class (they have the same form of the Ashtekar constraints for complex general relativity with cosmological constant) \cite{AshPRD.36.1587,Dadhich-8-3-002,beng261}, their evolution is trivial and no new constraints arise. Therefore, they constitute a set of seven first-class constraints, and since we have nine configuration variables $\tensor{A}{_a^i}$, the number of physical (complex) DOF per space point is two. Besides, one can introduce the diffeomorphism constraint defined by $\tilde{\mathcal{D}}_a:=\tilde{\mathcal{V}}_a+\tensor{A}{_a^i}\tilde{\mathcal{G}}_i$, and the action (\ref{Kras2}) takes the equivalent form

\begin{eqnarray}
& S[\tensor{A}{_a^i},\tensor{\tilde{\Pi}}{^a_i},\tensor{A}{_0^i},N^a,\utilde{N}]=\nonumber\\
&\int_{\mathbb{R}}dt\int_{\Omega}d^3x\left(\tensor{\tilde{\Pi}}{^a_i}\tensor{\dot{A}}{_a^i}+\lambda^i\tilde{\mathcal{G}}_i+N^a\tilde{\mathcal{D}}_a+\utilde{N}\tilde{\tilde{\mathcal{H}}}\right),\label{Kras3}
\end{eqnarray}

\noindent where $\lambda^i:=\tensor{A}{_0^i}-\tensor{A}{_a^i}N^a$. This implies that the Hamiltonian is given by

\begin{equation}\label{Hamil}
H=-\int_{\Omega}d^3x\left(\lambda^i\tilde{\mathcal{G}}_i+N^a\tilde{\mathcal{D}}_a+\utilde{N}\tilde{\tilde{\mathcal{H}}}\right),
\end{equation}

\noindent which is a linear combination of the constraints and therefore vanishes on shell. Thus, we have shown that the Hamiltonian formulation of Krasnov's action leads to the same phase space of the Ashtekar formalism. 

\section{Conclusion}

By starting from the Plebanski formulation (\ref{plebanAct}) for general relativity with cosmological constant, we have obtained its equivalent pure connection formulation (Krasnov's action) (\ref{krasnaction}) in a systematic fashion. Although it depends on a particular square root of $\tilde{M}$, the resulting equations of motion can be shown to imply Plebanski's equations regardless of the chosen root~\cite{krasnprl}, and so it describes complex general relativity with a nonvanishing cosmological constant. A pure connection formulation for vanishing cosmological constant is still lacking.

We also performed the Hamiltonian analysis {\it à la} Dirac of Krasnov's action, and we found that the phase space agrees with that of the Ashtekar formalism, as expected. Notice that a Hamiltonian analysis of a linearized version of (\ref{krasnaction}) was given in Ref.~\cite{Krasn84.024034} (see also~\cite{Delf118}), but no strict Hamiltonian analysis of (\ref{krasnaction}) had been performed before.

The procedure followed here could be applied to the real BF formulations of general relativity~\cite{MM81.044033,MM85.064011} to find pure connection formulations of them. It would be very interesting to see the role played by the Immirzi parameter in those formulations if such formulations really existed. Work in this direction is in progress.

\section*{Acknowledgments}

We thank Riccardo Capovilla and José A. Zapata for their fruitful comments. This work was supported in part by Consejo Nacional de Ciencia y Tecnolog\'ia  (CONACyT), M\'exico, Grant No. 167477-F.

\bibliography{references}

\end{document}